\documentclass[9pt,twocolumn,twoside]{osajnl}

\journal{ol} 

\setboolean{shortarticle}{true} 

\title{Directional scattering from particles under evanescent wave illumination: the role of reactive power}

\author[1,*]{Lei Wei}
\author[1]{Michela F. Picardi}
\author[1]{Jack J. Kingsley-Smith}
\author[1]{Anatoly V. Zayats}
\author[1]{Francisco J. Rodr\'{\i}guez-Fortu\~no}

\affil[1]{Department of Physics, King's College London, Strand, London, WC2R 2LS, United Kingdom}
\affil[*]{Corresponding author: lei.wei@kcl.ac.uk}

\ociscodes{(290.0290) Scattering.}

\begin{abstract}
Study of photonic spin-orbital interactions, which involves control of the propagation and spatial distributions of light with the polarization of electromagnetic fields, is not only important at the fundamental level but also has significant implications for functional photonic applications that require active tuning of directional light propagation. Many of the experimental demonstrations have been attributed to the spin-momentum locking characteristic of evanescent waves. In this letter, we show another property of evanescent waves: the polarization dependent direction of the imaginary part of the Poynting vector, i.e. reactive power. Based on this property, we propose a simple and robust way to tune the directional far-field scattering from nanoparticles near a surface under evanescent wave illumination by controlling linear polarization and direction of the incident light.
\end{abstract}

\setboolean{displaycopyright}{true}
\begin{document}
\maketitle
Photonic spin-orbital interactions (SOI) \cite{Cardano2015, Bliokh2015,Lodahl2017} have been of great interest in recent years, and various experiments\cite{Rodriguez-fortuno2013,Lin2013, Petersen2014,OConnor2014} have been carried out to demonstrate the optical counterpart of the quantum spin-Hall effect, which seeks to control the propagation and spatial distribution of light with polarization. A particularly intriguing group of demonstrations\cite{Rodriguez-fortuno2013,Petersen2014,OConnor2014} can be attributed to the transverse spin and the inherent spin-momentum locking of evanescent waves\cite{Bliokh2014,Aiello2015,VanMechelen2016}, where the transverse electric or magnetic field spin is determined by the direction of propagation. The directional excitation of fields based on spin-momentum locking introduces a spin degree of freedom in designing novel nanophotonic devices such as nanorouters\cite{Petersen2014}, polarimeters\cite{Espinosa-soria2017}, and non-reciprocal optical components\cite{Sayrin2015}. This directional light-matter interaction is also considered to have an impact on quantum optics\cite{Lodahl2017} for the development of complex quantum networks. 

Additionally, there exist other mechanisms to control the directionality based on the interference between electric and magnetic dipoles. One of such concepts was proposed by Kerker et al. for hypothetical magnetic spheres\cite{Kerker1983} which got renewed interest following the experimental demonstrations of Huygens dipoles with high index dielectric nanoparticles fulfilling Kerker's condition for zero 'backward' scattering\cite{Geffrin2012,Person2013,Fu2013}. The Kerker's condition for zero 'forward' scattering is however considered to be hard to realize for natural material particles under single plane wave illumination\cite{Zambrana-Puyalto2013, Engheta2010}. While early demonstrations generate orthogonal in-phase electric and magnetic dipoles to achieve directional radiating wave propagation, the Kerker's condition has since been generalized\cite{Liu2018,Zambrana-Puyalto2013} not only to include higher order multipoles\cite{Liu2014, Alaee2015} but also extended for near field directionality\cite{Picardijanus2017}. Another type of dipolar source, the so-called Janus dipole where the two linear electric and magnetic dipole moments are in quadrature phase to each other, has been proposed to achieve distinctive near field directionality\cite{Picardijanus2017}. Sources combining spin and Huygens dipoles lead to angular tuning of the directionality\cite{Wei2017}. Excitations with structured fields\cite{Xi2016,Wei2017} are also proposed to achieve broadband active tuning of directional scattering of nanoantennas. The dipole moments excited by focused vector beams are, however, very sensitive to the position of the particle in the focal field\cite{Neugebauer2016}, which poses a very strict requirement for experimental realization. 

In this Letter, we propose to achieve switchable directional scattering by the use of a single evanescent wave as the excitation field. We exploit the fact that, apart from intrinsic spin-momentum locking, there exists another locked relation between the electric and magnetic field components of a wave based on the imaginary part of the Poynting vector, i.e. the reactive power. The complex Poynting vector of a monochromatic electromagnetic field is given by $\mathbf{S} = \mathbf{E} \times \mathbf{H}^*$. From this equation we see that the electric and magnetic fields must have orthogonal components in order to give rise to a power flow. As is well known, the real part of the complex Poynting vector determines the net flow of electromagnetic energy, averaged over time, while the imaginary part yields the reactive power, representing the amplitude and direction of oscillation of the instantaneous Poynting vector at each point, associated with no net power flow. Notice that the phase between the orthogonal electric and magnetic fields determines whether the Poynting vector will be real or imaginary. When $\mathbf{E}$ and $\mathbf{H}$ are orthogonal and in phase, there is a net power flow (real $\mathbf{S}$). When they are in quadrature phase, there is a reactive power (imaginary $\mathbf{S}$). A propagating plane wave has a purely real Poynting vector because $\mathbf{E}$ and $\mathbf{H}$ are in phase, while a standing wave with equal incident and reflected amplitudes has a purely imaginary Poynting vector, as $\mathbf{E}$ and $\mathbf{H}$ are in quadrature phase. In an evanescent wave, the Poynting vector has both a real component parallel to the direction of propagation and an imaginary component parallel to the direction of evanescent decay. However, an important property of this reactive power is often overlooked, as we shall show: the imaginary Poynting vector components of a transverse magnetic (TM) and a transverse electric (TE) evanescent wave are antiparallel, pointing towards or against the direction of decay depending on linear polarization of the excitation light. 

We can exploit this polarization-dependent reactive power of evanescent waves in order to achieve directional scattering. We propose a simple mechanism: the use of a particle with electric and magnetic polarizabilities with a quadrature phase difference between them. This particle will generate scattered $\mathbf{E}$ and $\mathbf{H}$ fields whose relative phase is retarded or advanced by $\pi/2$ with respect to the incident field. Intuitively, this particle is able to convert incident reactive power into scattered net power flow, and vice-versa. Since the evanescent wave has opposite reactive power directions for different polarizations, we are able to tune the unidirectional scattering of the particle to be along or opposing the direction of evanescent wave decay by simply changing the polarization of the excitation. This simple and robust tuning mechanism of  unidirectional scattering is numerically investigated for nanoparticles excited by evanescent waves under attenuated total internal reflection(ATIR) configurations. It is shown that this method applies to various types of nanoparticles with different relative amplitudes of the electric and magnetic dipole polarizabilities, by choosing the proper transverse wavevector of the evanescent waves.

\begin{figure}[htp]
\centering
\includegraphics[width=0.45\textwidth]{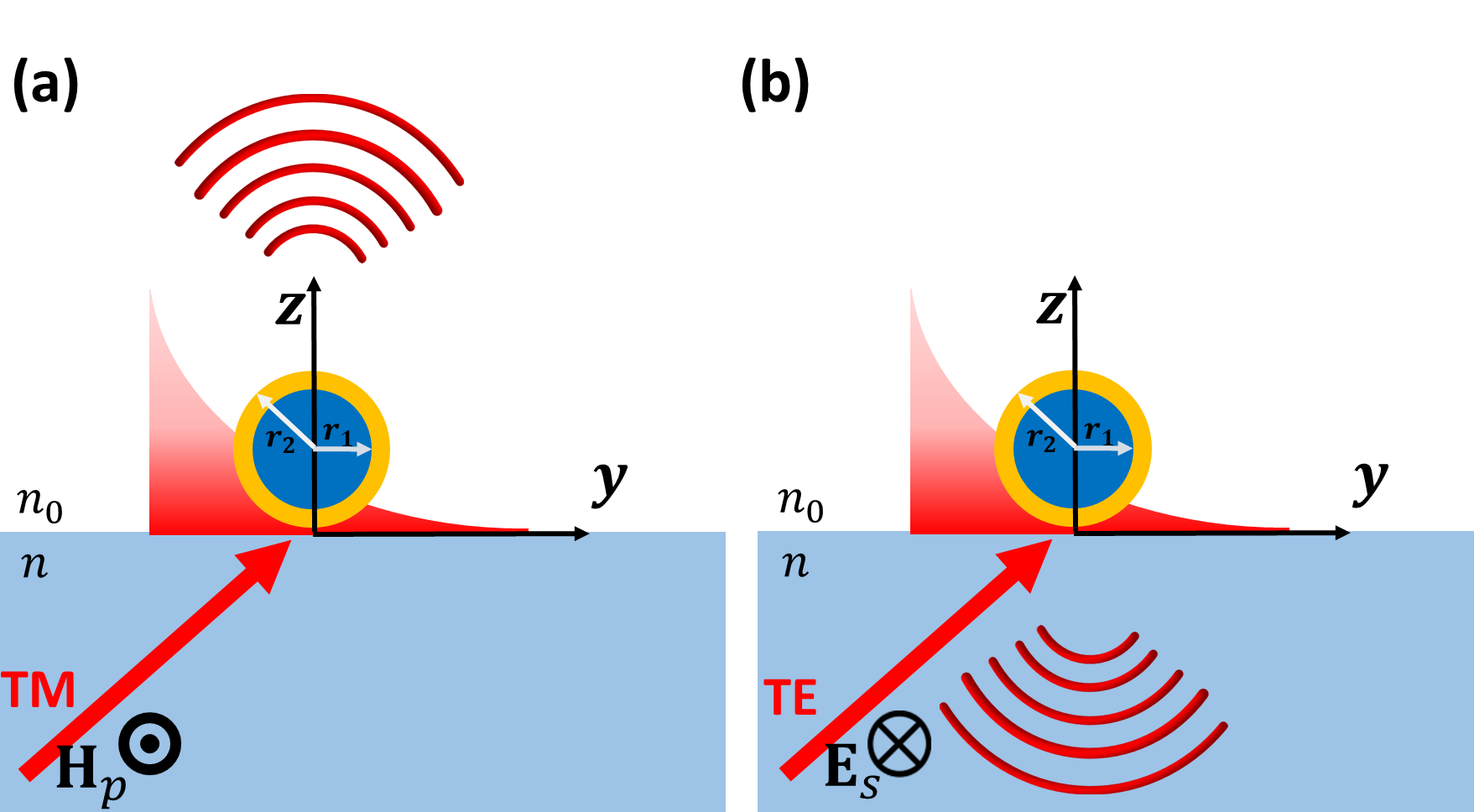}
\caption{A core-shell nanoparticle on an air-dielectric interface with a substrate of refractive index $n=2$ and illuminated by (a) TM polarized plane waves and (b) TE polarized plane waves incident from the substrate above the critical angle.}
\label{fig:config}
\end{figure}

An evanescent wave generated at an air-dielectric interface (Fig. \ref{fig:config}) has, in air, a complex wavevector $\mathbf{k}=\left (0,k_{\parallel}, i \gamma_z\right)$, where $\sqrt{k_{\parallel}^2-\gamma^2_z}=k_0$ and $k_0= 2\pi/\lambda$. Based on Maxwell equations, one can easily derive the fields for a TE polarized evanescent wave: 
\begin{align}
\mathbf{E}_s(\mathbf{r}) &=\left ( -E_s, 0, 0\right) \mathrm{exp}(ik_{\parallel}y-\gamma_z z),\nonumber\\
\mathbf{H}_s(\mathbf{r})&=\left(0, -\frac{i\gamma_z}{k_0}\frac{E_s}{Z_0}, \frac{k_{\parallel}}{k_0}\frac{E_s}{Z_0}\right) \mathrm{exp}(ik_{\parallel}y-\gamma_z z),  \label{eq:1}
\end{align}
where $Z_0$ is the impedance of free space and the fields of a TM polarized evanescent wave are
\begin{align}
\mathbf{E}_p(\mathbf{r})&=\left(0, -\frac{i\gamma_z}{k_0}E_p, \frac{k_{\parallel}}{k_0}E_p\right) \mathrm{exp}(ik_{\parallel}y-\gamma_z z),\nonumber\\
\mathbf{H}_p(\mathbf{r}) &=\left ( E_p/Z_0, 0, 0\right) \mathrm{exp}(ik_{\parallel}y-\gamma_z z).  \label{eq:2}
\end{align}

It follows from the transversality condition $\mathbf{k} \cdot \mathbf{E} = 0$ that the evanescent waves have transverse spin angular momenta and spin-momentum locking as an inherent property of evanescent waves \cite{Bliokh2014, VanMechelen2016}. Considering EM fields in Eq. (\ref{eq:1}) and Eq. (\ref{eq:2}), the complex Poynting vector $\mathbf{S} = \mathbf{E} \times \mathbf{H}^*$ of TE and TM polarized evanescent waves can be written as\\
\begin{align}
\mathbf{S}_s(\mathbf{r})&=\left[(-\frac{k_{\parallel}}{k_0}\frac{E_s^2}{Z_0})\mathbf{\hat{y}}-i(\frac{\gamma_z}{k_0}\frac{E_s^2}{Z_0})\mathbf{\hat{z}}\right] \mathrm{exp}(-2\gamma_z z),\nonumber\\
\mathbf{S}_p(\mathbf{r})&=\left[(-\frac{k_{\parallel}}{k_0}\frac{E_p^2}{Z_0})\mathbf{\hat{y}}+i(\frac{\gamma_z}{k_0}\frac{E_p^2}{Z_0})\mathbf{\hat{z}}\right] \mathrm{exp}(-2\gamma_z z).  \label{eq:Ssp}
\end{align}
As seen from Eq. (\ref{eq:Ssp}), both complex Poynting vectors are purely imaginary along the direction of evanescent decay $\hat{z}$ with a vital difference: the imaginary Poynting vector component of TE and TM polarized evanescent waves are opposite in sign. This can also be understood in a different way based on the complex Poynting theorem \cite{JacksonED} which reads $\nabla\cdot(\mathbf{E}\times\mathbf{H}^{*})=i\omega(\mathbf{B}\cdot\mathbf{H}^{*}-\mathbf{E}\cdot\mathbf{D}^{*})$. From Eq. (\ref{eq:Ssp}), we can see that the divergence of the complex Poynting vector of the evanescent waves $\nabla\cdot(\mathbf{E}\times\mathbf{H}^{*})=-2\gamma_z\hat{z}\cdot\mathbf{S}=-2\gamma_zS_z$ involves only $S_z$, so substituting this property into the complex Poynting theorem gives: 
\begin{equation}
S_z=-\frac{i\omega}{2\gamma_z}(\mathbf{B}\cdot\mathbf{H}^{*}-\mathbf{E}\cdot\mathbf{D}^{*})=-\frac{i\omega}{2\gamma_z}(\mu|\mathbf{H}|^2-\epsilon^{*}|\mathbf{E}|^2), \label{eq:3}
\end{equation}
so the sign of the imaginary vector component $S_z$ depends on whether the magnetic or the electric energy density dominates at each point in space, which happens in TE- and TM-polarized evanescent waves, respectively. In the remaining of this letter, we show how this often neglected but unique property could be used as a means to switch the unidirectional scattering of nanoparticles whose electric and magnetic polarizabilities are in quadrature phase.

Consider a nanoparticle having both an electric dipole polarizability $\alpha_{\mathrm{e}}=\frac{i6\pi\varepsilon_0}{k_0^3}a_1$ and a magnetic dipole polarizability $\alpha_{\mathrm{m}}=\frac{i6\pi}{k_0^3}b_1$ where $a_1$ and $b_1$ are the Mie coefficients of the electric and magnetic dipole modes and $\varepsilon_0$ is the vacuum permittivity. Nanoparticles with such properties have been demonstrated using high index dielectric materials like Si, Ge, etc. In the spectral region where the dipole modes dominate, the scattering of the nanoparticle is equivalent to the radiation of a source with an electric dipole moment $\mathbf{p}=\alpha_{\mathrm{e}}\mathbf{E}$ and a magnetic dipole moment $\mathbf{m}=\alpha_{\mathrm{m}}\mathbf{H}$, where $\mathbf{E}$ and $\mathbf{H}$ are the illuminating EM fields at the center of the nanoparticles. In the most general case, for any propagating or evanescent illumination with TE or TM polarization, we can state Kerker's condition as $|\Re\{\mathbf{p}/\mathbf{m}^*\}|=1/c_0$. When this condition is fulfilled, the scattering is completely suppressed along the direction of $-\Re\{\mathbf{p}\times\mathbf{m^*}\}$ \cite{Picardijanus2017,Nieto-Vesperinas2010}, which can be written in terms of the particle's polarizabilities as: $-\Re\{\alpha_{\mathrm{e}}\alpha_{\mathrm{m}}^*\}\Re\{\mathbf{E}\times\mathbf{H}^*\}-\Im\{\alpha_{\mathrm{e}}\alpha_{\mathrm{m}}^*\}\Im\{\mathbf{E}\times\mathbf{H}^*\}$. For a propagating wave, Kerker's condition is fulfilled for a particle with $a_1=b_1$, and the scattering is suppressed in the direction opposite to the real Poynting vector, resulting in zero backward scattering\cite{Fu2013,Person2013}. 

\begin{figure}[htp]
\centering
\includegraphics[width=0.5\textwidth]{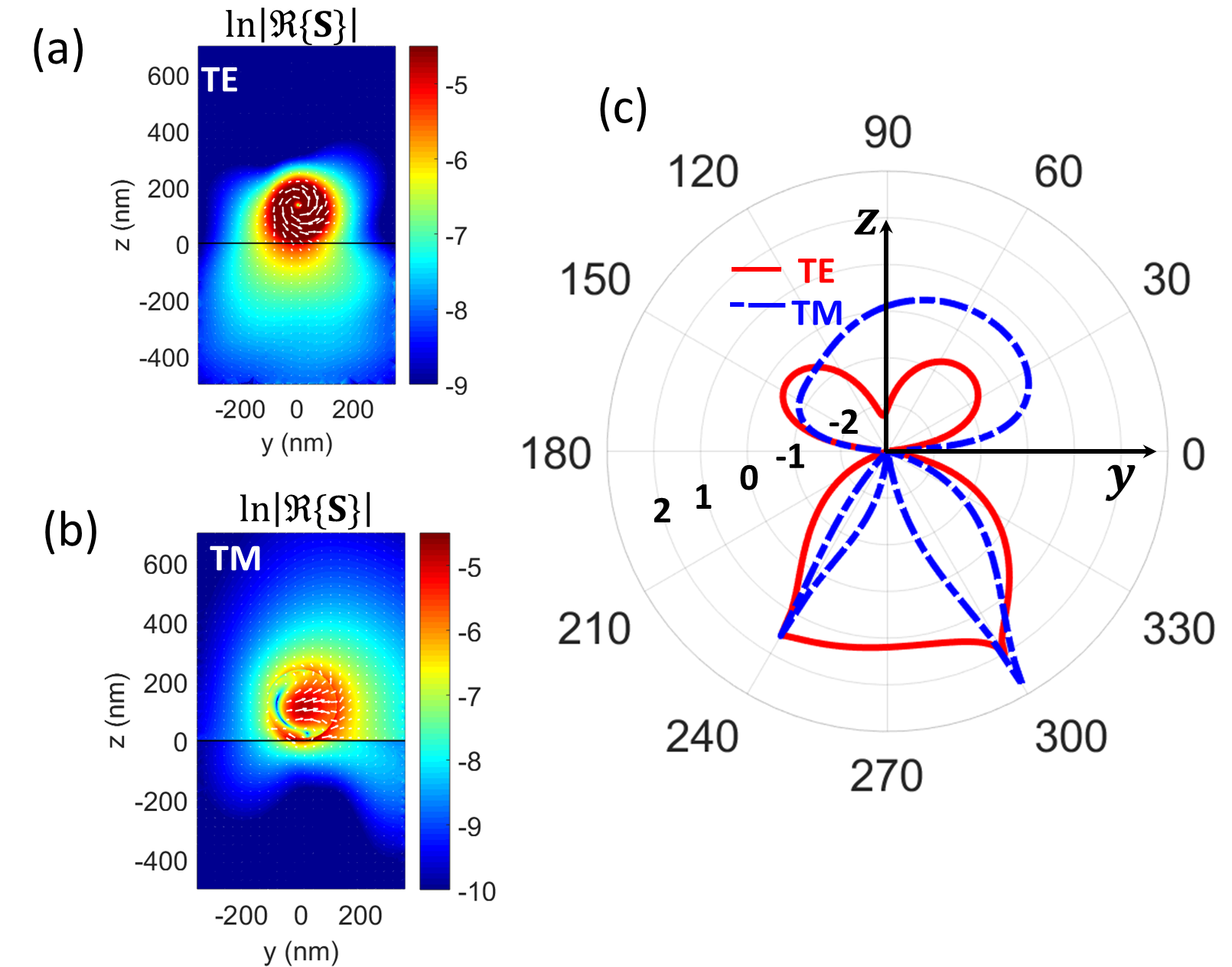}
\caption{The Poynting vector of the scattered field (total field with the particle minus the field of only air-dielectric interface) in the natural logarithm scale $\mathrm{ln}|\Re\{\mathbf{S}\}|$ of the core-shell nanoparticle with (a) a TE polarized and (b) a TM polarized light illumination with $k_{\parallel}=\sqrt{2}k_0$ from the substrate. The black line at $z=0$ nm in (a) and (b) denotes the air-dielectric interface. (c). The scattering patterns in the $yz$ plane with natural logarithm scale of the scattering power (in arbitrary units) of the core-shell nanoparticle with respectively TE and TM evanescent wave excitation.}
\label{fig:coreshell_scatter}
\end{figure}

With evanescent illumination, however, the direction of zero scattering also depends on the imaginary part of the Poynting vector. If a particle has its electric and magnetic polarizabilities in quadrature phase ($Arg\{b_1/a_1\}=\pi/2$), then its direction of zero scattering will be determined by the reactive power only. If such a particle is illuminated by a linearly polarized propagating wave, it will scatter like a Janus dipole\cite{Picardijanus2017}, showing unique directionality in the near-field but not in the far-field. However, under evanescent wave illumination, it will convert the reactive power in the near field into far field scattering. When Kerker's condition is fulfilled, this far field scattering can be zero along the direction $-\Re\{\mathbf{p}\times\mathbf{m^*}\}=-|\alpha_{\mathrm{e}}\alpha_{\mathrm{m}}^*|\Im\{\mathbf{E}\times\mathbf{H}^*\}$, i.e. opposite to the direction of the imaginary Poynting vector. As we have discussed previously, the sign of the imaginary Poynting vector component of an evanescent wave flips depending on whether it is TE or TM polarized. This unique property of an evanescent wave enables switching of the unidirectional scattering by simply changing its polarization. For TE polarized evanescent waves, when $\gamma_z/k_0=|a_1/b_1|$ so that $\mathrm{p}_x=\mathrm{m}_y/c_0$, there will be no scattering along the direction of $+\hat{z}$. For TM polarized evanescent waves, when $\gamma_z/k_0=|b_1/a_1|$ so that $\mathrm{m}_x/c_0=\mathrm{p}_y$, the scattering along the direction $-\hat{z}$ will be zero. The directionality of this type of particle under evanescent wave illumination is achieved parallel to the direction of decay, which is in stark contrast with the case of Kerker's condition on a propagating wave.\\
The concept of controlling the unidirectional scattering by polarization of the evanescent wave can be verified by a numerical study of the scattering of a nanoparticle on an air-dielectric interface as illustrated in Fig. \ref{fig:config}. The nanoparticle is illuminated by a linearly polarized evanescent wave. We will firstly study a specially designed core-shell particle composed of a high index dielectric core (inner radius $r_1=113$ nm and refractive index 3.7) and a metallic shell (outer radius $r_2=119$ nm and permittivity $\varepsilon_{\mathrm{m}}=-26.067 + 0.303i$ at the wavelength $\lambda=755$ nm). The parameters are chosen such that the scattering electric and magnetic dipole coefficients $a_1$ and $b_1$, calculated from the analytical Mie theory, have the relation $b_1\approx i a_1$ at $\lambda=755$ nm. The nanoparticle is excited by a linearly polarized evanescent wave with tangential wavevector $k_{\parallel}=\sqrt{2}k_0$ (such that $\gamma_z/k_0=1$) at the wavelength $\lambda=755$ nm. If we neglect the multiple scattering between the particle and a dielectric surface, $\mathrm{m}_x/c_0=\mathrm{p}_y$ for TE polarization and $\mathrm{p}_x=\mathrm{m}_y/c_0$ for TM polarization. The scattering patterns are calculated by Finite Element Method (FEM) and are shown in Fig. \ref{fig:coreshell_scatter}. By changing the polarization of the evanescent waves from TM to TE, the zero scattering direction can be switched from $-\hat{z}$ to $+\hat{z}$.

\begin{figure}[htp]
\centering
\includegraphics[width=0.5\textwidth]{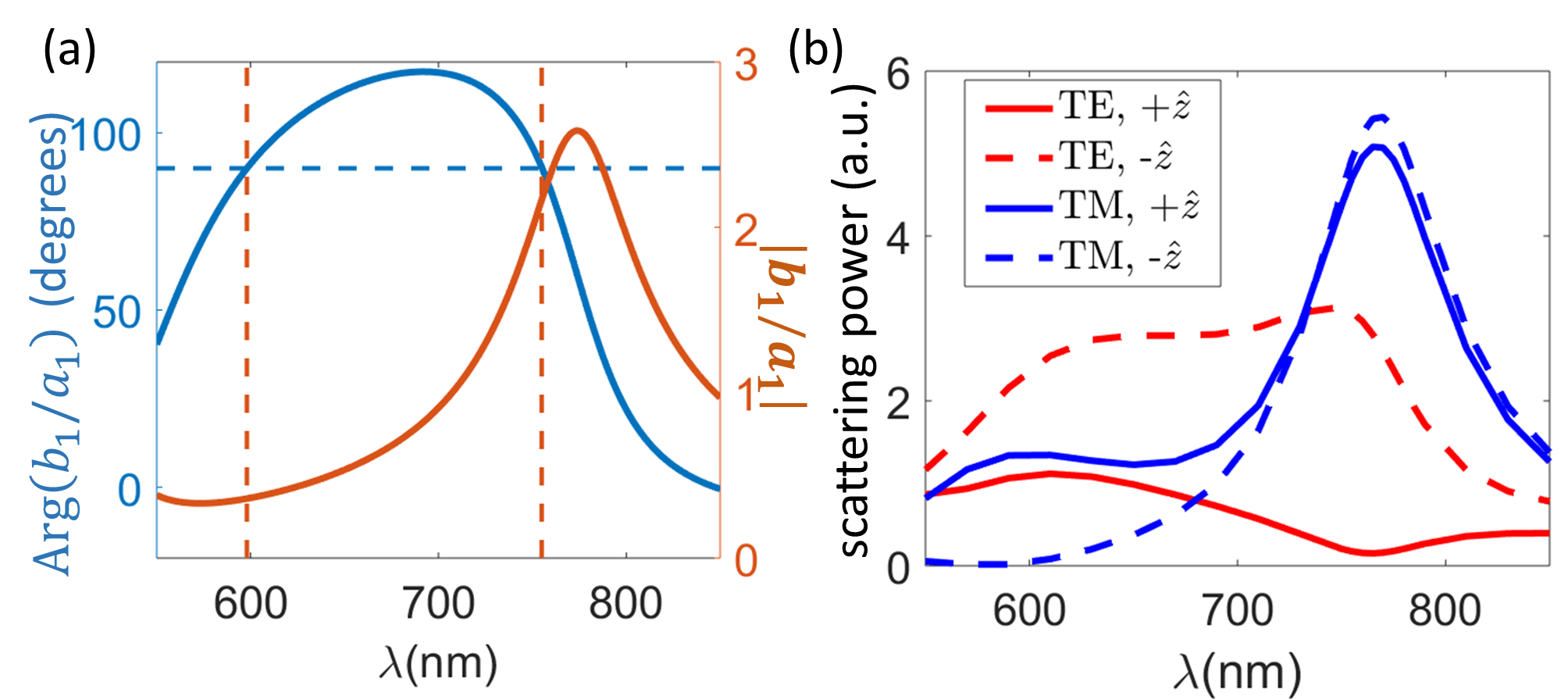}
\caption{(a). Amplitude and phase of the ratio of the magnetic and electric dipole coefficients $b_1/a_1$ of a dielectric nanosphere with refractive index 3.7 and radius 100 nm; (b). The scattering power spectrum along $\pm \hat{z}$ directions when the nanosphere is illuminated by TE and TM evanescent waves with $k_{\parallel}=2\sin(33.5^{\mathrm{o}})$.}
\label{fig:si_scatter}
\end{figure}

This unidirectional scattering switching by means of polarized evanescent wave excitation is, however, not restricted to the specially designed core-shell structure in which $b_1\approx i a_1$. We can also apply it to simpler high index dielectric nanospheres. For instance, the phase difference and amplitude ratio between the magnetic and electric dipole coefficients of a nanosphere with refractive index 3.7 and radius 100 nm is shown in Fig. \ref{fig:si_scatter}(a) within the spectral range where the higher order multipole modes are negligible. As can be seen, there are two wavelengths where the magnetic and electric dipoles have a $\pi/2$ phase difference: $b_1/a_1=2.166i$ at $\lambda=755$ nm and $b_1/a_1=0.3599i$ at $\lambda=598$ nm. To match Kerker's condition for the dipoles in the $xy$ plane, one can simply adjust the incidence angle. In this way, the transverse wave vector $k_{\parallel}$ can be selected, which will determine the relative amplitude between incident electric and magnetic fields in the evanescent wave, according to Eqs. (\ref{eq:1}) and (\ref{eq:2}). This ratio can be tuned such that $\gamma_z/k_0=|a_1/b_1|$ for TE and $\gamma_z/k_0=|b_1/a_1|$ for TM polarized evanescent waves. For example, in order to achieve the directional scattering along z direction at the wavelength $\lambda=755$ nm with a TE polarized evanescent wave excitation, we need  $k_{\parallel}/k_0=\sqrt{1+|a_1/b_1|^2}=1.10$ while for a TM polarized evanescent wave excitation we need $k_{\parallel}/k_0=\sqrt{1+|b_1/a_1|^2}=2.39$. Similarly, at the wavelength $\lambda=598$ nm, to fulfill the Kerker's condition along z direction, a transverse wavevector $k_{\parallel}/k_0=\sqrt{1+|a_1/b_1|^2}=2.95$ is required for a TE polarized evanescent wave while $k_{\parallel}/k_0=\sqrt{1+|b_1/a_1|^2}=1.06$ for a TM polarized evanescent wave excitation. In Fig. \ref{fig:si_scatter}(b), the spectral dependence of the scattering power of the nanoparticle along $\pm\hat{z}$ directions are calculated by FEM simulations. The incident light comes from the dielectric substrate at the angle $33.5^{\mathrm{o}}$ which generates TE and TM evanescent wave excitations for the particle with $k_{\parallel}=2\sin(33.5^{\mathrm{o}})=1.10$. This wavevector matches roughly the Kerker's condition of the dipole components in the $xy$ plane for the particle excited by TE evanescent waves at $\lambda=755$ nm and TM evanescent waves at $\lambda=598$ nm. Indeed, we can see from Fig. \ref{fig:si_scatter}(b) that the directional scattering is achieved with minimum scattering direction along $+\hat{z}$ for TE polarized excitation at $\lambda=755$ nm and along $-\hat{z}$ for TM excitation at $\lambda=598$ nm. This again verifies that when the Kerker's condition is fulfilled, the preferred scattering direction is opposite to the reactive power direction of TE and TM evanescent wave excitations. 

Another interesting observation from Fig. \ref{fig:si_scatter} is that $\lambda=755$ nm, where $\mathrm{Arg}(b_1/a_1)=\pi/2$, is close to the magnetic dipole (MD) resonance. This can be explained by the fact that the phase of the magnetic dipole polarizability is changing rapidly around the MD resonance while the phase of the electric dipole polarizability is changing slowly, and thus $\mathrm{Arg}(b_1/a_1)$ passes through $\pi/2$ near the MD resonance. As the wavevector $k_{\parallel}=2.39$ needed to achieve unidirectional scattering with TM polarized evanescent wave at this wavelength is much higher than the one applied $k_{\parallel}/k_0\approx1.10$ in Fig. \ref{fig:si_scatter}(b), the particle's magnetic dipole moment dominates over the electric one. This leads to a large scattering power contrast between TE and TM excitations along the $+\hat{z}$ direction, which may inspire applications like polarization dependent filters, switching, etc. \\

Comparing our setup with the realization of Huygens dipoles by exciting high index dielectric nanoparticles with propagating waves\cite{Fu2013,Person2013}, the proposed realization of Huygens dipoles using evanescent waves has several unique features that makes it stand out. Existing demonstrations of Kerker's condition with propagating wave excitation\cite{Fu2013,Person2013} are only possible to achieve zero backward scattering with dual nanoparticles that have $a_1=b_1$. Due to the fact that the electric and magnetic field vectors of propagating waves are locked to the propagation wavevector, and it is hard to find a dielectric nanoparticle that has strictly anti-dual properties $a_1=-b_1$\cite{Zambrana-Puyalto2013}, there aren't many possibilities to tune the scattering direction with a single free propagating wave. However, as shown in this letter, it is possible to switch the unidirectional scattering not along the direction of wave propagation but along the direction of decay of the evanescent wave illuminations.
Using the difference in sign between the reactive power in TE and TM evanescent waves, we have shown unidirectional scattering of a particle with $\mathrm{Arg}(b_1/a_1)=\pi/2$, either composite or dielectric, can be achieved by choosing the proper $k_{\parallel}$, and the scattering direction can be switched by simply changing the polarization of the evanescent wave illumination. It is also interesting to notice that the condition $a_1=b_1$ of the propagating wave realization of Kerker's condition often occurs at a far off-resonance wavelength where both $|a_1|$ and $|b_1|$ are relatively weak. However, the requirement $\mathrm{Arg}(b_1/a_1)=\pi/2$ in our proposed setup, for instance the high index particles in Fig. \ref{fig:si_scatter}, often happens in a spectral range very close to the electric or magnetic dipole resonances where abrupt phase change occurs. Thus the scattered power tends to be large. In this Letter, we showed the usefulness of the polarization-dependent direction of reactive power in evanescent waves, and we proposed a realization to control the directional scattering using polarized evanescent waves, which can inspire applications for directional light coupling, displays, polarization dependent filtering, switching and forces, amongst many other.

This work was supported by European Research Council Starting Grant ERC-2016-STG-714151-PSINFONI and EPSRC (UK). A.Z. acknowledges support from the Royal Society and the Wolfson Foundation.

\end{document}